\documentclass[10pt]{article}
\usepackage{amssymb}
\usepackage{amsthm}
\usepackage{authblk}
\usepackage{setspace}
\usepackage{ wasysym }
\usepackage{caption}
\usepackage[textwidth=6.5in,textheight=8.25in,centering]{geometry}
\usepackage[pdftex]{graphicx}

\onehalfspace
\begin{document}
\begin{center}
{\LARGE{\textbf{In-situ spectroscopic studies of viologen based electrochromic device}}}

\vspace{0.5 cm}
\textit{ Suryakant Mishra, Haardik Pandey, Priyanka Yogi, Shailendra K. Saxena, Swaroop Roy, Pankaj R. Sagdeo and Rajesh Kumar}\footnote{Corresponding author email: rajeshkumar@iiti.ac.in}

\vspace{0.5 cm}
Material Research Laboratory, Discipline of Physics \& MSEG, Indian Institute of Technology Indore, Simrol-453552, Madhya Pradesh, India

\vspace{1 cm}
ABSTRACT

\end{center}
Fabrication and operation of simple solid state electrochromic devices using ethyl viologen diperchlorate in a polymer matrix is presented here. In-situ Raman and transmission/absorption studies have been done to establish the origin of bias induced color change, between a transparent and navy blue color, in the electrochromic device. The origin of bias induced color change has been attributed to the bias induced redox switching between its viologen dication and free redicle forms. Fundamental reason behind colour changes of viologen molecule has been established. In-situ UV-Vis spectra reveals that the navy blue color of the device under biased condition is not due to increase in the transparency corresponding to blue wavelength but due to suppression of the transparency corresponding to the complementary colors. Absorption modulation has been reported from the device with good ON/OFF contrast of the device.
\vspace{0.5cm}

\textbf{Keywords:}  Raman spectroscopy, UV-Vis spectroscopy, Viologen, Electrochromism
\begin{figure}[h]
\begin{center}
\includegraphics[width=8cm]{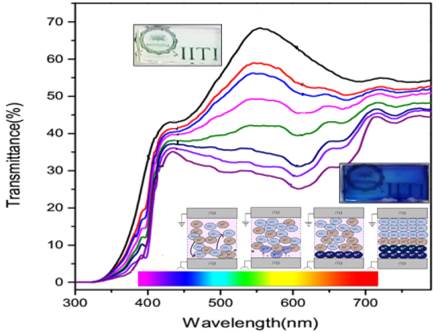}
\captionsetup{labelformat=empty}
\caption{-----------------------xxxxxxxxxxxxxxxxx-----------------------}
\end{center}
\end{figure}

\newpage
\section{Introduction}
An electrochromic device, as name suggests, is a device which changes color as a result of an electrical bias. This phenomenon of bias induced color change, called electrochromism, was first observed in the nineteenth century[1]. Electrochromism has been studied in recent times because of many applications in science and technology.[2–-5] Detailed reviews are available in the literature about electrochromic effects from different chemicals and materials.[2,6] Electrochromic effects are observed as a result of redox processes either in a solid-state device or in an electrochemical cell. Origin of this property lies in the fact that many materials show multiple redox states with different optical properties (e.g. absorption spectra).  Different electrochromic materials are used as the active material, which could be in the form of metal oxide like tungsten oxide[7] or a cationic molecule.[8–-10] 

One particularly intriguing class of materials in this context is bipyridinium species, which are formed upon N,N-diquaternization of 4,4'- bipyridine, also known as viologens. An example of such a violgen molecule is shown in scheme-1. Viologen is an organic compound for potential use in flexible electrochromic devices due to its activity in reversible redox reaction and excelent electron accepting nature. In other words, Viologens[9],[11–-14] are reducible materials used in solid-state electrochromic devices, which may undergo one- or two-electron reductions.[9],[11–-14] Ease of processing and use have stimulated intensive research into the electrochromic properties of viologens. The current research was undertaken to better understand viologen electrochromism and to simplify the device design for realization of a solid-state electrochromic device.
\begin{figure}[h]
\begin{center}
\includegraphics[width=5cm]{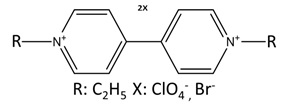}
\captionsetup{labelformat=empty}
\caption{Scheme 1: Ethyl Viologen molecule}
\end{center}
\end{figure}
 
 Controlling the often complex stoichiometry of metal oxides is a complication when fabricating electrochromic devices based on these materials[15,16]. In addition, many metal oxides lack sufficient transparency for use in a wide range of applications.  In contrast, viologen based materials are solution processable and can yield flexible electrochromic devices. In order to achieve an electrochromic response in the solid state, an ionically conducting matrix is required as well as a redox counter-reaction. 
 
 Very recently, lithium perchlorate in polyethylene oxide (PEO) has been used successfully as a solid electrolyte for organic electronics in general and thin film transistors in particular [10],[17-–21]. Ethyl viologen in PEO matrix can be a good choice for making viologen based eletrochromic devices, at least for testing purposes. Ethyl viologen diperchlorate (EV (ClO$_4$)$_2$) can be a good candidate because it contains EV$^{2+}$ which is a good electron acceptor and also contains perchlorate ion which can act as electrolyte.  Its solution in an appropriate solvent is transparent and does not absorb visible light. After accepting an electron it gets reduced and form radical cation EV$^{\newmoon +  }$.   In the reduced state it shows absorption peak at 396 and 606nm which indicate blue color of the device[22]. The advantage with viologen is that it shows reversible switching between the two oxidation states mentioned above. In other words, EV$^{\newmoon +  }$ can go back to EV$^{2+}$ by oxidation, which shows no absorption in the visible region which is transparent state of the device. Details of redox reactions taking place during oxidation and reduction of EV have been provided in SI separately. A solid state electrochromic device can be made using viologens by appropriately controlling the redox process using an electric bias.
\begin{figure}
\begin{center}
\includegraphics[width=10cm]{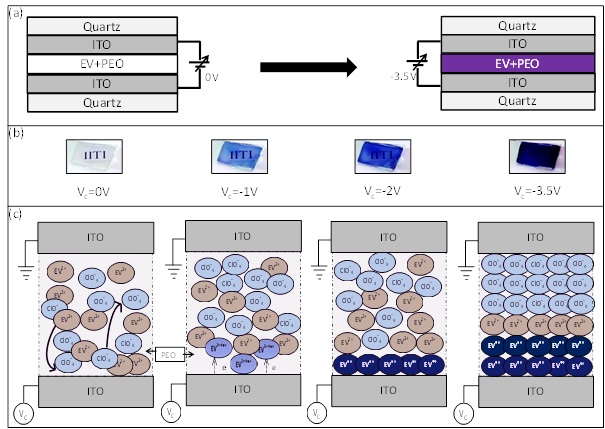}
\captionsetup{labelformat=empty}
\caption{Figure 1: (a) Schematic representation of cross-bar geometry device connecting through the external biasing arrangement (b) photographs of device without and with applied bias, (c) Schematic model of change in oxidation state of viologen inside the device as a result of applied bias.}
\end{center}
\end{figure}
 Many of the viologen applications are largely based on its ability to change in the electron absorption spectrum by reversible oxidation–reduction reaction. This feature of viologen can be tuned by altering the substituents on the nitrogen and used it for applications spanning from photochemistry and solar energy conversion, molecular wire and switches. Recently viologen has been used as single molecular transistor Viologen is used as chemical quencher for improvement of quantum yield of nanoemitter[23]. Viologens are frequently used as electrochemical label, photochemical probe or in studies on the electron-transport processes or oxidative damage in DNA[24].
 
 In the present work a simple electrochromic device has been fabricated in two geometries to study the fundamental reason of colour change. A clear understanding may enable one to get other colors than blue. In depth studies have been done by observing bias induced changes in EV species using spectroscopic techniques in-situ. The in-situ experiments helps in understanding the underlying mechanism for bias induced color change which is required for making a versatile electrochromic device. Nondestructive spectroscopic techniques like Raman and UV-Vis may prove to be handy in analyzing the mechanism for the same. The two geometries used to understand the bias induced color change enables us to carry out in-situ measurements for an in-depth analysis. In the simplest geometry, the EV (in an appropriate matrix) has been sandwiched between two Indium Tin Oxide (ITO) electrodes on quartz substrate to get the device in cross-bar geometry (CBG). Similar device was fabricated by drop casting EV in between two gold electrodes separated by few microns where both the electrodes are on the same plane. Schematic to show both the geometries are shown in Fig. S1 in the supplementary information (SI). OFG device is used for in-situ Raman spectroscopy with electrical circuit arrangement to see the molecular change by applying bias. Raman spectroscopy experiments in OFG device and result clearly revealed connection between electrochemical doping and Raman peak position. Both UV-Vis and Raman spectroscopic data reconcile the correlation between charge carriers and color changes in viologen base devices. The free radical cation has been depicted as the fundamental charged species responsible for the color changing properties in ethyl viologen. Appropriate control experiments have been carried out to validate the conclusions. 
 
\begin{figure}
\begin{center}
\includegraphics[width=10cm]{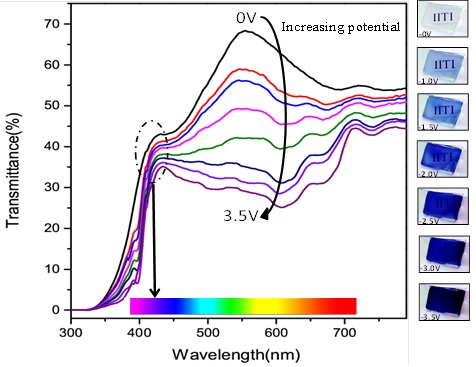}
\captionsetup{labelformat=empty}
\caption{Figure 2: In-situ UV-Vis spectroscopy to show changes in transmittance of the CFG device by applying bias in the range of 0 to -3.5V. Inset showing colour appearance of the device at various potential.}
\end{center}
\end{figure}

\section{Experimental Details}
Commercially available chemicals from Alfa Aesar and Sigma Aldrich were used for preparing samples in the present study. Polyethylene oxide (PEO, Alfa Aesar, MW = 100 000), Ethyl viologen diperchlorate (98\%, Sigma Aldrich), Sodium borohydride (NaBH$_4$, Alfa Aesar), and Acetonitrile (ACN, anhydrous, 99.8\%, Sigma Aldrich) were used as received.
Active material used for fabricating electrochromic devices was prepared in the solution form by mixing 4 wt.\% ethyl viologen diperchlorate [EV (ClO$_4$)$_2$] in acetonitrile and 5 wt.\% PEO in acetonitrile. The PEO solution was filtered through a 0.45 $\mu$m PTFE filter before adding the viologen solution. The solution (PEO+EV) has been deposited over the electrode by dropping 3 $\mu$l of solution in such a way that the drop centers on the E$_1$-E$_2$ gap in the OFG (Fig. S1 in SI) and then dried the film in vacuum.

Cross-bar geometry (CBG) device has been fabricated by placing a film of EV+PEO between two transparent conducting electrodes of ITO deposited on quartz substrate using magnetron sputtering deposition [25]. The EV+PEO film was obtained on an ITO coated substrate using spin coating. After spin coating, second ITO electrode was laminated face to face on the spin coated substrate. The extreme part of both the electrodes was painted by silver glue for making connection with external power supply. UV-Vis spectroscopy was performed using Cary 60 UV-Vis spectrophotometer of Agilent whereas Raman spectroscopy was done using LABRAM HR spectrometer using a 633 nm excitation source.
\begin{figure}
\begin{center}
\includegraphics[width=10cm]{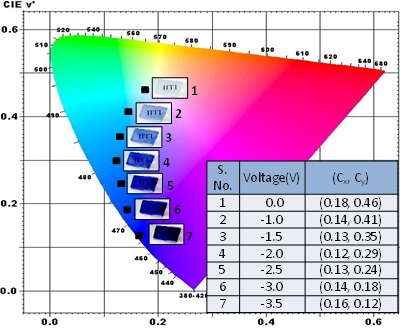}
\captionsetup{labelformat=empty}
\caption{Figure 3: CIE chart fitting of different intensity of a colour for device. Inset table is for the colour coordinate with respect to applied bias.}
\end{center}
\end{figure}

\section{Results and Discussion}
Figure 1 is the combined pictorial representation of internal and external mechanism of CBG device. Fig. 1(a) shows cross sectional schematic view of CBG device with its biasing arrangement. Fig.1(b) is corresponding actual photographs of CBG device on various voltages. Fig 1(c) corresponds to proposed model which shows molecular dynamics and redox reaction within the device as a result of applied bias. Fig.1(a) shows that when external potential is 0 V, device is in its OFF state which corresponds to the transparent state and thus `IITI' can be seen through the device clearly in Fig. 1(b). The transparency of the complete device reduces and becomes opaque after the application of -3.5 V bias as shown in Fig.1(b). Color change in viologen is a well reported phenomenon[26] and is often assigned to the different absorption/transmission properties exhibited by viologen species in its different oxidation states. Typical optical properties of chemically reduced viologen show that blue color originates due to the presence of EV$^{\newmoon +  }$  species  which can be obtained by chemically reducing EV$^{2+}$ (present in ethyl viologen diperchlorate) which is transparent. Alternatively, similar redox change can be induced in the material in the form of a solid state device as a result of electrical bias which provides the basis for an electronic electrochromic device.

\begin{figure}
\begin{center}
\includegraphics[width=10cm]{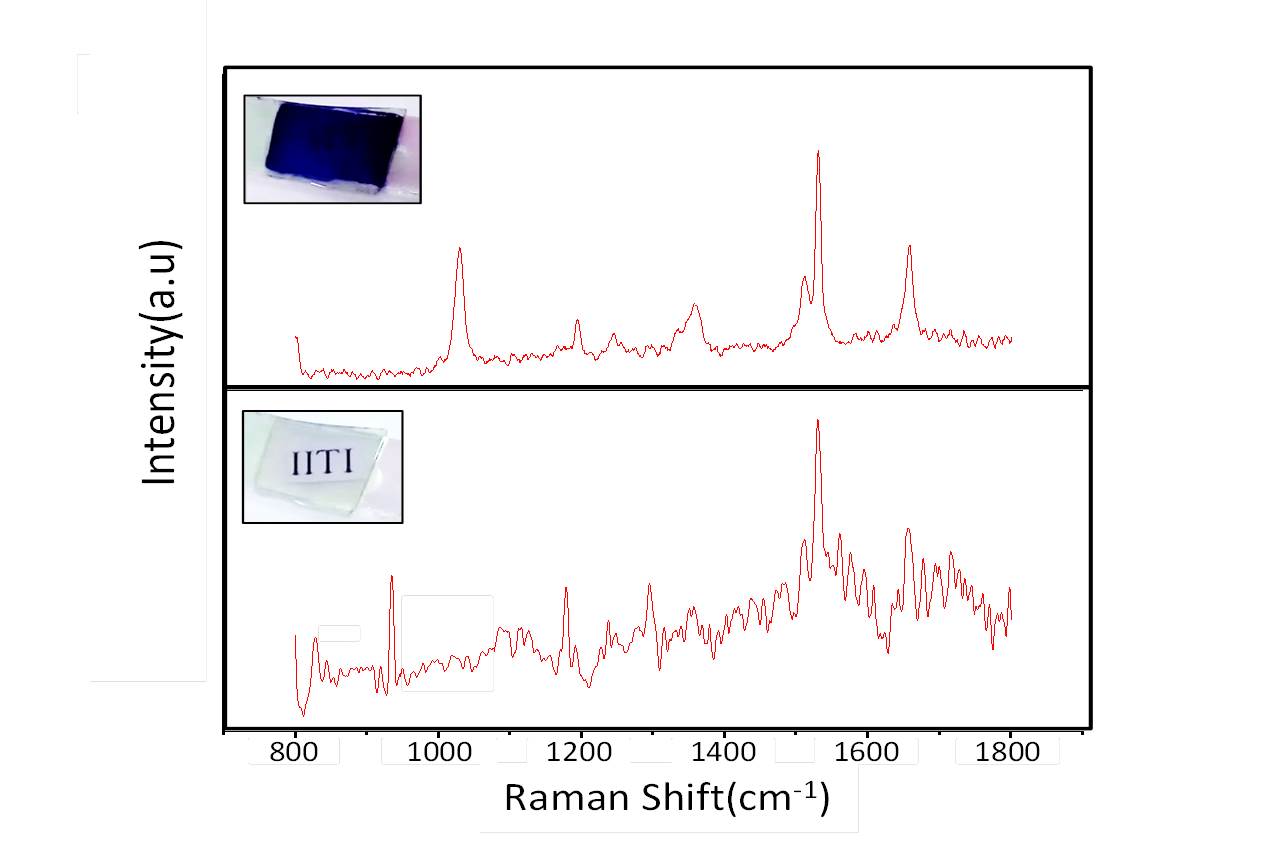}
\captionsetup{labelformat=empty}
\caption{Figure 4: In-situ Raman spectra in ON and OFF state of device in cross-bar geometry. Insets show actual photographs of the device in its ON and OFF states.}
\end{center}
\end{figure}

Mechanism responsible for bias induced color change as shown in Fig1 (b) can be understood as follows.  When potential is applied, electrons are supplied by the power supply and infusion of electron take place into the compound which may be reducing the viologen molecules near the -ve electrode to get the opaque (ON) state when most of the viologen molecule gets reduced. EV$^{2+}$ and ClO$^{4-}$ species (in PEO matrix) are randomly distributed in between the two ITO electrodes in the absence of bias (V=0). When bias is applied (V = -1V) electrons supplied by the –ve electrode is accepted by EV$^{2+}$ to get reduced to EV$^{\newmoon +  }$  to initiate the color change. As higher voltages are applied more reduction take place and large number of EV$^{\newmoon +  }$  gets generated which results in opaque device. In addition, the perchlorate and EV$^{\newmoon +  }$ ions move towards positive and negative electrodes respectively. As a result, complete opaque nature of device is obtained with arrangement of ions in the device as shown in the case of V = -3.5V bias case in Fig. 1(c).

Alternatively, it may also be possible that bias induced color change is associated with the (electric) field induced structural change[27] of the molecule which may lead to band gap variation and hence change in optical properties[28]. To discount this effect control experiment was done by sandwiching viologen between two quartz substrates without the ITO electrode. Schematic for this experiment has been shown in Fig S2 in the SI.  The present biasing condition does not provide electrons to the molecule (for possible reduction) but do provide field for any field-related structural change. No color change was observed upto a bias voltage of 50V which suggests that the observed color switching is not related to field induced structural change and more likely reason being the transmission change due to redox process.

The blue color visible as a result of bias in Fig. 1 has only qualitative meaning and needs more scientific approach to identify the presence of different wavelength components in the transmitted light when the device is seen in a visible light. Perceived color by human eye must be cross checked with scientific results (with optical tools) before making any conclusions. To understand the color changing phenomena in CBG device the best suitable technique is Ultra-violet visible (UV-Vis) spectroscopy. From UV-Vis spectra one can understand the actual wavelength of light which is more affected by the reduction of viologen molecule.
\begin{figure}
\begin{center}
\includegraphics[width=10cm]{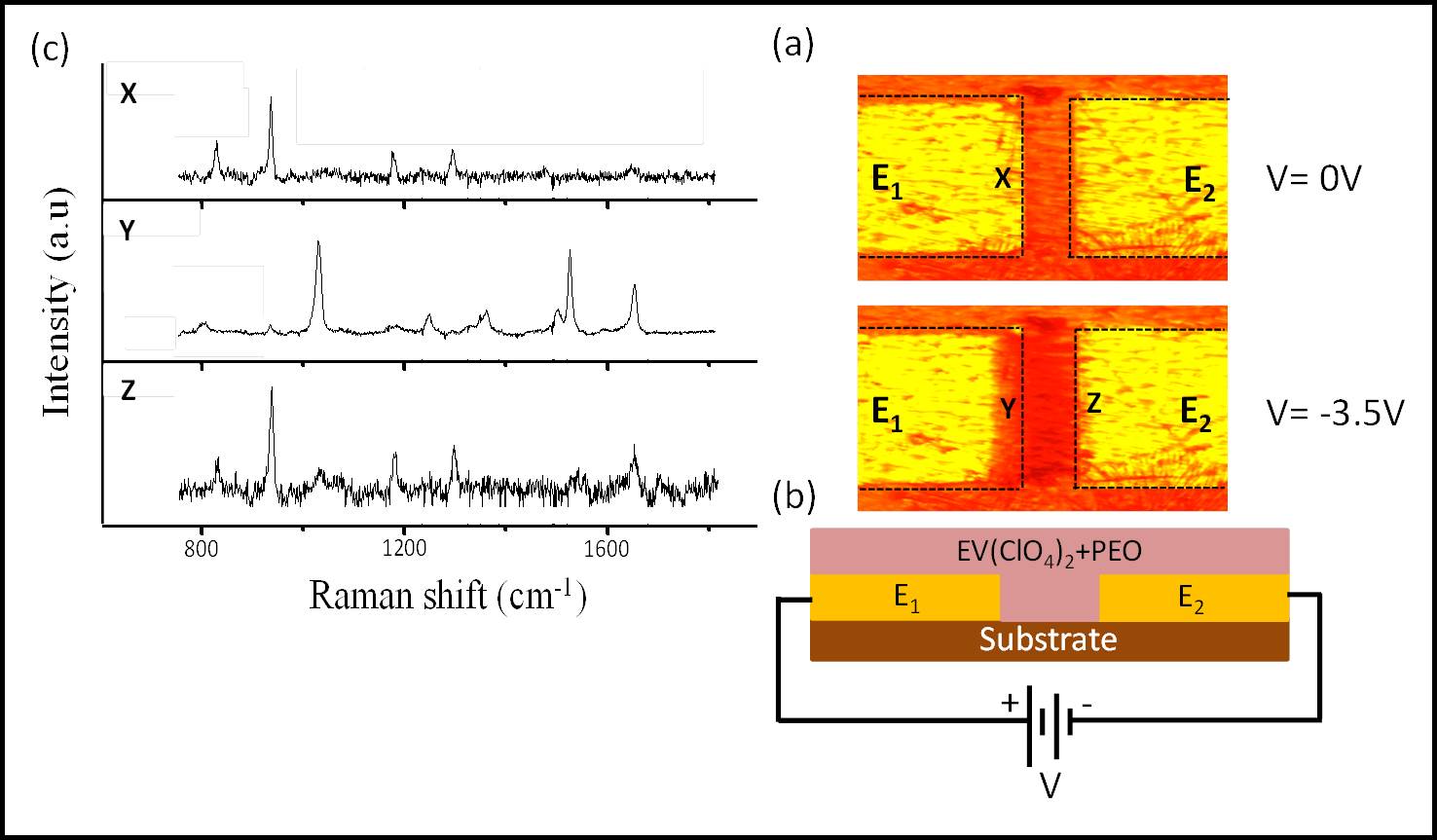}
\captionsetup{labelformat=empty}
\caption{Figure 5: (a) Optical image of OFG with and without bias. (b) Schematic cross section view of OFG device with biasing arrangement. (c)In-situ Raman spectra recorded on the various position of the OFG device.}
\end{center}
\end{figure}
In-situ UV-Vis spectroscopy has been performed to investigate optical property of the device which are clearly visible from naked eyes. Figure 2 represent transmission spectra with respect to change in bias voltage using UV-Vis spectroscopy. Transmittance (\%) of the film depends upon the band-gap of the material so the band-gap modulation is carried out by the application of voltage During in-situ measurement, bias was increased gradually upto -3.5V(in steps of 0.5 V) and transmittance of the device is measured in the range between 300 to 800 nm. The transmission spectrum under zero bias condition has been recorded as reference data, to see any variation in it after bias application. Under no bias condition, only ITO and EV$^{2+}$ (in PEO matrix) species are present and a peak appears around 560 nm. The peak around 560 nm gets suppressed as the biasing potential is increased and finally it changed into a valley shaped curve for a voltage of -3.5 V. It can be noticed from Fig.2 that there is not much change in \%T at the wavelengths of blue (425 nm) and red (725 nm) colours as compared to the wavelengths in between these two. The device exhibited a minimum transmittance in the visible region at around 590 (orange-yellow) nm with a transmittance of 22 \% at -3.5V. As a result, the counter colour of orange-yellow, i.e, blue, is observable to the visible eyes. This also means that blue color of the device in ON state is not due to the enhancement of blue wavelength but due to suppression of other colors in the visible range. 

It is clearly observed from the transmittance spectrum corresponding to -3.5V that transmission is suppressed in the range of 500 to 700 nm. Also, from the color wheel we can say the counter colour of most suppressed colour in transmission spectra will be the colour of the film[29]. The simple analysis of UV-Vis spectra allows one to understand the origin of the color appearance of the electrochromic device. This also may enable one to understand that \% T corresponding to a given wavelength may be tuned by appropriately enhancing/suppressing the \%T spectra at other wavelengths. Relative \%T corresponding to different wavelengths decide the exact colour of appearance and can be obtained from CIE (Commission Internationale de l'Eclairge International Commission on Illumination) color calculator (Osram Sylvania[30]).

Figure 3 shows the CIE color calculator chart with the photograph of our actual devices with different applied bias. The CIE color calculator has been used here to calculate CIE coordinates corresponding to the device appearance color for a given applied bias. Initially at zero bias (OFF) device is transparent hence it allow the white light to pass through it[31]. As the applied bias voltage increase, only blue color spectrum pass through it as compared to green and red. According to colour appearance,  related point is observe on different coordinate of the chart[32] as shown in Fig. 3. Finally it gets saturated at dark blue colour and their corresponding coordinate on CIE chart are (0.16, 0.12). Hence except blue color, device absorbs all colors in its ON state. The CIE chart is very useful as it says that if somehow the material, to be used in device, is tuned in such a way that the color corresponding to coordinates (0.5,0.5) is allowed to transmit from the device a red color appearance will be observed. If one identifies the species responsible for appearance of one particular color corresponding to a given coordinate, the same can be used to make a devise giving that particular color. Before implementing this idea, it is very important to understand the species responsible for a particular color's appearance. This investigation can be done using Raman scattering experiments.
\begin{figure}
\begin{center}
\includegraphics[width=10cm]{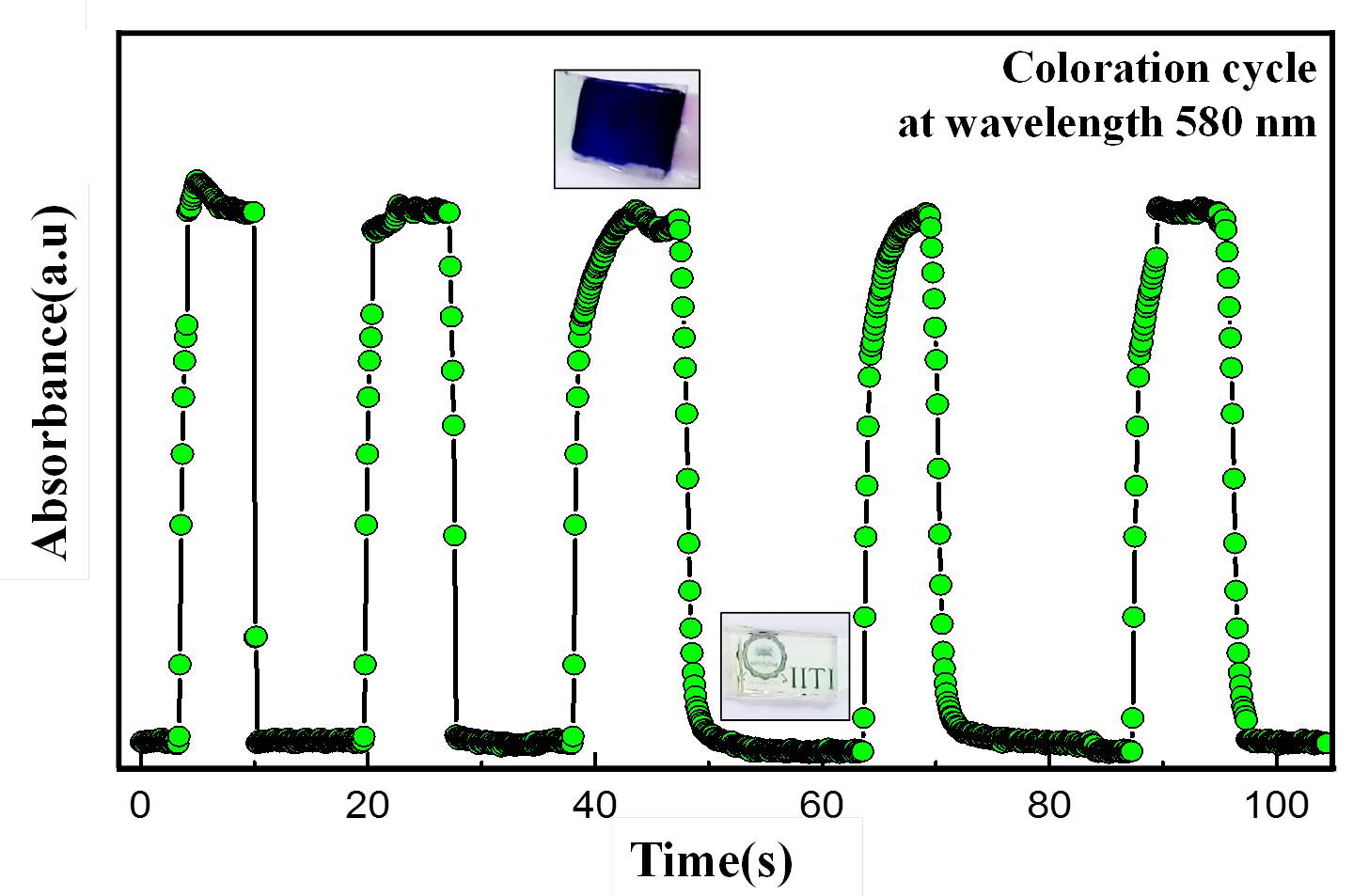}
\captionsetup{labelformat=empty}
\caption{Figure 6: In-situ UV-Vis spectroscopy coloration and de-coloration cycle or absorbance change by biasing ON and OFF state within a time period. Inset image for static analysis of change in transmittance.  }
\end{center}
\end{figure}
Figure 4 shows Raman spectra of the device in its OFF and ON states. Raman spectrum of the device in its OFF state is similar to the  EV(ClO$_4$)$_2$+PEO mixture in powder form. This means that the device contains the EV$^{2+}$ form of viologen in the OFF state. Raman spectrum of the device in the ON state is different from its OFF state with some new peaks appearing and some disappearing. This certainly means that the species present in the ON and OFF states, giving rise to the Raman peaks, are different. It is also likely that the peaks appearing in the ON state are obtained as a result of reduction of the species present in the OFF state. To confirm this Raman spectra in Fig. 4 have been compared with the Raman spectra from EV$^{2+}$ and EV$^{\newmoon +  }$  as obtained by chemically reducing the EV(ClO$_4$)$_2$ using NaBH$_4$ which is shown in Fig. S3 in the SI. Comparison of Fig. 4 and Fig. S3 shows that the OFF device contains EV$^{2+}$ whereas the ON device contains EV$^{\newmoon +  }$  in majority which means that EV$^{2+}$ reduces to EV$^{\newmoon +  }$  as a result of applied bias more likely at the –ve electrode. This cannot be confirmed in the CBG because a spatial Raman spectroscopy is not possible.

To confirm this, in-situ Raman spectroscopy of the device in OFG has been carried out.  Device in OFG was prepared by drop casting the active material (EV+PEO) on a pre-fabricated electrode system where two gold electrodes are a few microns apart. A drop of prepared sample in the gap completes electrical circuit and device will start working like electrochemical cell. Top view of the finished device is shown in Fig. 5(a) without and with bias applied. The polarity scheme of the power supply to give bias is shown in Fig. 5(b), which shows the cross-sectional schematic of the device. A color change on electrode E1 as a result of –ve bias on this electrode can be seen which is similar to the bias induced colour change in Fig. 2 . This can be appreciated by comparing point X and Y in Fig. 5(a). Raman spectra has been recorded from points X,Y and Z (Fig. 5(a)) to compare the effect of bias on the present chemical species in the device. These points have been chosen for Raman measurements so that a comparison can be done between Raman signal from a virgin device (point X), negative terminal (point Y, where reduction is possible) and positive terminal (point Z) under biased condition. Fig. 5(c) shows Raman spectra at different position of OFG device under different bias condition. Microscopic image in Fig.5(b) of OFG device shows colour change only on the negative electrode of the device. During the electrochemically reduction, intensity of color between E$_1$-E$_2$ and on E$_1$ electrode changed which attributes reduction of EV$^{2+}$ molecule.

The corresponding Raman scattering studies have been done on E$_1$ and E$_2$. Alphabets X, Y, Z on the microscopic images indicate the positions from where the Raman signals were collected. Raman spectrum in figure 5(c) X was recorded from the E$_1$ terminal of of an unbiased device, which shows major peak at 932 cm$^{2+}$, the peaks corresponding to EV$^{2+}$. Raman spectra 5(c)Y and 5(c)Z  have been recorded under bias of -3.5V on E$_1$ and E$_2$ respectively. Some additional peaks appear on the E$_1$ electrode in biased condition, spectrum 5(c)Y shows two prominent peaks corresponding to reduced viologen which are at 1028 and 1528 cm$^{-1}$. With biasing potential, peak intensity is increases as the applied potential increases which shows the formation of more free radical cations (EV$^{\newmoon +  }$ )There is no change in the spectrum recorded on E$_2$ where +ve terminal of the power supply is connected. 

For the comparative study of various peaks positions, Raman spectra of neutral and chemically reduced form of EV(ClO$_4$)$_2$ has been carried out to verify spectra obtain from electrode E$_1$ and E$_2$ under different bias conditions (Fig. 5(c)). To identify and verify[33] the changes in EV$^{2+}$, EV$^{\newmoon +  }$ and EV$^{o}$ concentration Raman peaks of chemically reduced state of EV were studied. The solid EV(ClO$_4$)$_2$  powder was reduced using an aqueous solution of sodium borohydride (NaBH$_4$).  Raman spectrum from this reduced state of EV(ClO$_4$)$_2$ is shown in Fig S3 in SI and it was observed that the EV powder turns dark-bluish immediately after mixing with NaBH$_4$ due to reduction of EV. This change of colour is due to variation in absorption spectrum. The same solution turns back into transparent due to oxidation of EV$^{\newmoon +  }$ to EV$^{2+}$ as higher oxidation state is the more stable state in the ambient conditions. Raman spectrum from chemically reduced form of EV(ClO$_4$)$_2$ shows peaks at 1028 and 1528 cm$^{-1}$ which are signatures of EV$^{\newmoon +  }$ species.(These peaks are marked by stars in figure S4). It is now safe to conclude that the blue color observed in Fig. 2 as a result of applied bias is due to the EV$^{\newmoon +  }$ species being generated at the –ve terminal as a result of reduction of EV$^{2+}$. By combining the data in Figs 2, 5 \& 3S, one can see that the Raman spectrum from the dark region on the device (E$_1$ electrode in Fig. 5(a)) shows the presence of EV$^{\newmoon +  }$ species whereas the virgin device shows the signature of stable EV$^{2+}$. Therefore, the Raman spectra permit the conclusion that the mechanism of bias induced color change in the electrochromic device is due to formation of free radical cation from viologen dication.

Above discussion confirms the origin of color change in the electrochromic device. It will be interesting to see how fast and how many times the electrochromic device be turned ON and OFF. Figure 6 shows ON and OFF cycles in terms of normalized absorbance corresponding to 580 nm which has been studies within a specific time period[34]. Initially at 580 nm, on zero bias absorbance is in its lowest value, Al which reaches the highest value Ah by applying potential (-3.5V) making it opaque for this wavelength. Figure 6 shows that device can be turned ON and OFF five times in $\sim$100 seconds. It is worth mentioning that it takes more time in returning to OFF state after removal of the applied bias because the returning to OFF state is natural and is not under any external perturbation. A 50 \% transparency in the second cycle onwards has been considered as OFF state whereas ON state corresponds to 100 \% transparency. Additionally, contrast between the ON and OFF states can be calculated by defining colour contrast ratio (CCR) using the following equation[35-–37].
$$CCR(\%) = \frac{T_i-T_f}{T_i}\times 100 $$

Here T$_i$ and T$_f$   are the initial and final values of transmittance for the device at wavelengths 580 nm. The values of T$_i$ and T$_f$ are 80\% and 12\% giving a 68\% variation in transmittance and CCR of nearly 85\%. Figure 6 also shows that it takes approximately 1.5 seconds to switch between the ON and OFF states of the device. An electrochromic device with 85\% CCR is a very good device with an ability to switch in 1.5 seconds which can be improved if an external method is used for turning OFF the device of course by modifying the device composition and geometry appropriately.

\section{Conclusions}

In-situ spectroscopic studies of viologen based electrochromic devices reveal that bias induced color change appears due to bias induced change in oxidation state of viologen. The blue color of the device as a result of negative bias is obtained due to presence of viologen frees redical (EV$^{\newmoon +  }$). Raman spectra from the devices under this bias condition confirms the presence of viologen free redical in both the geometries of device. Raman signature corresponding to the free redical is seen on the electrode connected to the negative polarity of the power supply (used to bias the device) whereas EV$^{2+}$ (viologen dication) is present on the other electrode which was connected to the positive polarity. Species on the positive terminal of the device is same as the species present in the whole device in its virgin state. Since the cross bar geometry device is symmetric, it shows color switching between transparent to navy blue aster application of bias irrespective of the polarity. In-situ UV-Vis spectra reveals that the navy blue color of the device under biased condition is not due to increase in the transparency corresponding to blue wavelength but due to suppression of the transparency corresponding to the complementary colors. A 85 \% contrast and 68 \% variation in color has been observed in the device

\subsection*{Acknowledgements}
Authors acknowledge financial support from Department of Science and Technology (DST), Govt. of India. Authors thank Professor R.L. McCreery (NINT, University of Alberta, Canada) for providing gold electrodes and Professor A. Subrahmanyam (IIT Madras, India) for providing ITO electrodes. Authors are thankful to Dr. V. Sathe (UGC-DAE CSR, Indore, India) for Raman measurements. Authors are also thankful to MHRD, Govt. of India for providing fellowships.

\end{document}